\documentclass[%
 reprint,
groupedaddress,
showpacs,
 amsmath,amssymb,
 aps,
prb,
]{revtex4-1}

\usepackage{graphicx}
\usepackage{dcolumn}
\usepackage{bm}

\begin{document}

\newcommand{\be}{\begin{equation}}
\newcommand{\ee}[1]{\label{#1}\end{equation}}
\newcommand{\bem}{\begin{eqnarray}}
\newcommand{\eem}[1]{\label{#1}\end{eqnarray}}
\newcommand{\eq}[1]{Eq.~(\ref{#1})}
\newcommand{\Eq}[1]{Equation~(\ref{#1})}
\newcommand{\vp}[2]{[\mathbf{#1} \times \mathbf{#2}]}


\title{Reply to Comment on ``Symmetry of Kelvin-wave dynamics and the Kelvin-wave cascade in the $T=0$ superfluid turbulence'' [arXiv:1208.4593]}

\author{E.  B. Sonin}
 \affiliation{Racah Institute of Physics, Hebrew University of
Jerusalem, Givat Ram, Jerusalem 91904, Israel}

\date{\today}

\begin{abstract}
I discuss the Comment  by L'vov and Nazarenko (arXiv:1208.4593) aiming at refutation of my perviously published criticism of their mechanism of the Kelvin-wave cascade. It is important that in their Comment L'vov and Nazarenko admitted that the Hamiltonian, from which they derived their mechanism, is not tilt-invariant. This provides  full ammunition to their critics, who believe that their mechanism is in conflict with tilt symmetry of the Kelvin-wave dynamics in an isotropic space.
\end{abstract}

\maketitle


In their Comment \citet{LNcom} defend their mechanism of the Kelvin-wave cascade against the criticism presented in my recent paper. \cite{SonKW} First I would like to focus on the central question to the L'vov-Nazarenko mechanism, to which they  have finally  replied  in their Comment.

The original Hamiltonians in single vortex line dynamics (either governed by the Bio-Savart law, or  in the local-induction approximation) are definitely tilt-symmetric, i.e. are invariant with respect to line tilting or rotation of the coordinate frame with respect to the line. Meanwhile, the mechanism of  L'vov and Nazarenko\cite{Lau,Lvo} depends  on the tilting angle   and therefore is not tilt-invariant.\cite{Koz10,KozC}  This manifests itself in linear dependence of the effective 6-waves interaction amplitude from  one of  the  wave  numbers $k_1$   in the infrared limit $k_1 \to 0$. As a result of it, the assumption of locality for the six-wave interaction fails, invalidating the  mechanism of the Kelvin-wave cascade   previously suggested by \citet{Koz}.

In principle, it is quite common that some process or state has symmetry lower than that of the Hamiltonian. It is called broken symmetry. The dispute is whether broken tilt symmetry takes place in the Kelvin-wave cascade. The problem is complicated by the fact that the Kelvin-wave dynamics requires a weak-perturbation expansion in a series, and any truncation of the series strictly speaking leads to violation of the tilt symmetry. Nevertheless, according to  the paper   \cite{SonKW}  it does not compromise tilt symmetry because the latter is broken only in terms of high order, which are beyond accuracy of the perturbation theory. In particular, in the theory  of the Kelvin-wave cascade, which requires terms up to the 6th order, a proper expansion violates tilt symmetry only in terms of the 7th order or higher. So a proper truncation of the series provides tilt symmetry with sufficient accuracy. The problem with Hamiltonian used by L'vov  and Nazarenko in their calculations was that tilt symmetry is broken in essential terms (of the 6th order or less). It is important that in the Comment L'vov and Nazarenko  finally have accepted this. But (the first argument of L'vov  and Nazarenko):

{\em 1. L'vov  and Nazarenko do not see anything wrong in the fact that their Hamiltonian violates tilt symmetry.}

So they do not see any problem with the fact that their approximate Hamiltonian does  not respect tilt symmetry while the exact Hamiltonian does! This makes a dispute about broken symmetry irrelevant, because this is not a case of  broken tilt symmetry but of {\em absent} tilt symmetry. Definitely one may agree that the mechanism of L'vov  and Nazarenko is legitimate for such a Hamiltonian and their claim that ``the presence of tilt symmetry does not imply the absence of linear  infrared asymptotics in the nonlinear interaction coefficients'' is trivially correct, simply because something (tilt symmetry), which is absent, cannot imply anything. However, as a next step L'vov  and Nazarenko must admit that their calculation is detached from the original problem of Kelvin-wave dynamics in an isotropic space. L'vov  and Nazarenko time and again invite their opponents to  ``put the Hamiltonian on the table'' or detect possible error in their cumbersome calculation. This request can be returned back to them: ``Put on the table the Hamiltonian respecting tilt symmetry.'' Otherwise their calculation has no credit independently of whether it was algebraically correct or not. 

 L'vov  and Nazarenko refer to \citet{LLN} who showed (quotation from the Comment) ``that the tilt symmetry does not prevent the interaction coefficients of {\em any} order to have linear in $k$ asymptotics.'' Nobody (not the paper  \cite{SonKW} in any case) argues against this statement. But the question is whether a Hamiltonian, or a combination of these coefficients, which determines a physical effect,  may depend on the line tilt (i.e., linear in $k$ asymptotics).

{\em 2. According to L'vov  and Nazarenko, it is impossible to eliminate their mechanism by rotation of the coordinate axis.} 

They state that such a rotation can remove an axis tilt (to which intensity of their mechanism is proportional) only for a single infrared mode, whereas in an ensemble of such modes  one cannot eliminate a tilt in all these modes simultaneously. This argument is an attempt to jump from a simpler single-infrared-mode problem to a more complex one (ensemble of modes)  without any justification.  Note that their mechanism (if correct) must be fully valid for a single infrared mode, when it can be eliminated by rotation of the axis.  
This is exactly what the paper had in mind. In fact, their derivation was done for a single mode in the infrared limit $k_1\to 0$. What happen with this mechanism in an ensemble of infrared modes is an interesting question only if the mechanism does not vanish in a single mode. But if it does vanish there, 
 averaging of zeros cannot provide anything else than zero.

{\em 3. L'vov  and Nazarenko criticize the statement that tilting of the axis affects distribution in $k$
space but not frequencies of Kelvin modes.}

They call the statement incorrect and then explain: ``The nonlinear frequency shift has nothing to do with any sort of redistribution in the $ k$-space but rather it is a change of frequency, at a fixed $k$, with respect to the frequency of the linear waves.''
I fully agree with this since it is exactly what was said in my paper. But  the question is whether tilting of the axis changes frequency at a {\em fixed} $k$ or  the wave number $k$ at a {\em fixed} frequency $\omega $. The paper\cite{SonKW} presents an elementary transformation of  the axis rotation demonstrating that the frequency but not the wave number is fixed. It is a direct consequence of the fact that rotation of axes in space, which does not involve time, cannot change the period of oscillation.
As far as  L'vov  and Nazarenko do not challenge this simple transformation their objection is not justified.

The next criticism addresses the qualitative discussion of the strong Kelvin-wave turbulence when the perturbation-theory parameter is not small:

{\em 4. L'vov  and Nazarenko qualify this discussion as incorrect but still eventually producing the correct answer.}

I   think that they judge this discussion from the position of high standards, which this hand-waving discussion did not pretend. I  agree that the series expansion itself becomes a risky idea for strong turbulence, but still can provide a hint to real processes. I  agree that my analysis totally ignores the question of locality, but remind that their belief of  nonloclality of terms of the 6th order is not accepted by everyone and in fact is criticized (see above). In all, I have presented a cartoon picture of a complicated effect indeed. But their harsh criticism of it is the same as blaming of a cartoonist for distortion of an image of a real personage, which is shown in the cartoon.


In summary, accepting that the Hamiltonian, which they use in their theory  of the Kelvin-wave cascade, is not tilt invariant,  L'vov  and Nazarenko fully confirm that their mechanism is not relevant for the Kelvin-wave dynamics in our isotropic space. All other issues disputed by them are next in importance.

\begin{acknowledgments}
I thank 
Victor L'vov, Sergey Nazarenko, and Sergey Nemirovskii 
for interesting discussions.
The work was supported by the grant of the Israel Academy of Sciences and Humanities.
\end{acknowledgments}


\begin{thebibliography}{8}%
\makeatletter
\providecommand \@ifxundefined [1]{%
 \@ifx{#1\undefined}
}%
\providecommand \@ifnum [1]{%
 \ifnum #1\expandafter \@firstoftwo
 \else \expandafter \@secondoftwo
 \fi
}%
\providecommand \@ifx [1]{%
 \ifx #1\expandafter \@firstoftwo
 \else \expandafter \@secondoftwo
 \fi
}%
\providecommand \natexlab [1]{#1}%
\providecommand \enquote  [1]{``#1''}%
\providecommand \bibnamefont  [1]{#1}%
\providecommand \bibfnamefont [1]{#1}%
\providecommand \citenamefont [1]{#1}%
\providecommand \href@noop [0]{\@secondoftwo}%
\providecommand \href [0]{\begingroup \@sanitize@url \@href}%
\providecommand \@href[1]{\@@startlink{#1}\@@href}%
\providecommand \@@href[1]{\endgroup#1\@@endlink}%
\providecommand \@sanitize@url [0]{\catcode `\\12\catcode `\$12\catcode
  `\&12\catcode `\#12\catcode `\^12\catcode `\_12\catcode `\%12\relax}%
\providecommand \@@startlink[1]{}%
\providecommand \@@endlink[0]{}%
\providecommand \url  [0]{\begingroup\@sanitize@url \@url }%
\providecommand \@url [1]{\endgroup\@href {#1}{\urlprefix }}%
\providecommand \urlprefix  [0]{URL }%
\providecommand \Eprint [0]{\href }%
\providecommand \doibase [0]{http://dx.doi.org/}%
\providecommand \selectlanguage [0]{\@gobble}%
\providecommand \bibinfo  [0]{\@secondoftwo}%
\providecommand \bibfield  [0]{\@secondoftwo}%
\providecommand \translation [1]{[#1]}%
\providecommand \BibitemOpen [0]{}%
\providecommand \bibitemStop [0]{}%
\providecommand \bibitemNoStop [0]{.\EOS\space}%
\providecommand \EOS [0]{\spacefactor3000\relax}%
\providecommand \BibitemShut  [1]{\csname bibitem#1\endcsname}%
\let\auto@bib@innerbib\@empty
\bibitem [{\citenamefont {L'vov}\ and\ \citenamefont
  {Nazarenko}(2012)}]{LNcom}%
  \BibitemOpen
  \bibfield  {author} {\bibinfo {author} {\bibfnamefont {V.~S.}\ \bibnamefont
  {L'vov}}\ and\ \bibinfo {author} {\bibfnamefont {S.~V.}\ \bibnamefont
  {Nazarenko}},\ }\href@noop {}arXiv:1208.4593 
  \BibitemShut
  {NoStop}%
\bibitem [{\citenamefont {Sonin}(2012)}]{SonKW}%
  \BibitemOpen
  \bibfield  {author} {\bibinfo {author} {\bibfnamefont {E.~B.}\ \bibnamefont
  {Sonin}},\ }\href@noop {} {\bibfield  {journal} {\bibinfo  {journal} {Phys.
  Rev. B}\ }\textbf {\bibinfo {volume} {85}},\ \bibinfo {pages} {104516}
  (\bibinfo {year} {2012})}\BibitemShut {NoStop}%
\bibitem [{\citenamefont {Laurie}\ \emph {et~al.}(2010)\citenamefont {Laurie},
  \citenamefont {L'vov}, \citenamefont {Nazarenko},\ and\ \citenamefont
  {Rudenko}}]{Lau}%
  \BibitemOpen
  \bibfield  {author} {\bibinfo {author} {\bibfnamefont {J.}~\bibnamefont
  {Laurie}}, \bibinfo {author} {\bibfnamefont {V.~S.}\ \bibnamefont {L'vov}},
  \bibinfo {author} {\bibfnamefont {S.}~\bibnamefont {Nazarenko}}, \ and\
  \bibinfo {author} {\bibfnamefont {O.}~\bibnamefont {Rudenko}},\ }\href@noop
  {} {\bibfield  {journal} {\bibinfo  {journal} {Phys. Rev. B}\ }\textbf
  {\bibinfo {volume} {81}},\ \bibinfo {pages} {104526} (\bibinfo {year}
  {2010})}\BibitemShut {NoStop}%
\bibitem [{\citenamefont {L'vov}\ and\ \citenamefont {Nazarenko}(2010)}]{Lvo}%
  \BibitemOpen
  \bibfield  {author} {\bibinfo {author} {\bibfnamefont {V.~S.}\ \bibnamefont
  {L'vov}}\ and\ \bibinfo {author} {\bibfnamefont {S.}~\bibnamefont
  {Nazarenko}},\ }\href@noop {} {\bibfield  {journal} {\bibinfo  {journal}
  {Pis'ma Zh. Eksp. Teor. Fiz.}\ }\textbf {\bibinfo {volume} {91}},\ \bibinfo
  {pages} {464} (\bibinfo {year} {2010})},\ \bibinfo {note} {[JETP Lett. {\bf
  91}, 428 (2010)]}\BibitemShut {NoStop}%
\bibitem [{\citenamefont {Kozik}\ and\ \citenamefont
  {Svistunov}(2010{\natexlab{a}})}]{Koz10}%
  \BibitemOpen
  \bibfield  {author} {\bibinfo {author} {\bibfnamefont {E.}~\bibnamefont
  {Kozik}}\ and\ \bibinfo {author} {\bibfnamefont {B.}~\bibnamefont
  {Svistunov}},\ }\href@noop {} {\bibfield  {journal} {\bibinfo  {journal}
  {Phys. Rev. B}\ }\textbf {\bibinfo {volume} {82}},\ \bibinfo {pages}
  {140510(R)} (\bibinfo {year} {2010}{\natexlab{a}})}\BibitemShut {NoStop}%
\bibitem [{\citenamefont {Kozik}\ and\ \citenamefont
  {Svistunov}(2010{\natexlab{b}})}]{KozC}%
  \BibitemOpen
  \bibfield  {author} {\bibinfo {author} {\bibfnamefont {E.~V.}\ \bibnamefont
  {Kozik}}\ and\ \bibinfo {author} {\bibfnamefont {B.~V.}\ \bibnamefont
  {Svistunov}},\ }\href@noop {} {\bibfield  {journal} {\bibinfo  {journal} {J
  .Low Temp. Phys.}\ }\textbf {\bibinfo {volume} {161}},\ \bibinfo {pages}
  {603} (\bibinfo {year} {2010}{\natexlab{b}})}\BibitemShut {NoStop}%
\bibitem [{\citenamefont {Kozik}\ and\ \citenamefont {Svistunov}(2004)}]{Koz}%
  \BibitemOpen
  \bibfield  {author} {\bibinfo {author} {\bibfnamefont {E.}~\bibnamefont
  {Kozik}}\ and\ \bibinfo {author} {\bibfnamefont {B.}~\bibnamefont
  {Svistunov}},\ }\href@noop {} {\bibfield  {journal} {\bibinfo  {journal}
  {Phys. Rev. Lett.}\ }\textbf {\bibinfo {volume} {92}},\ \bibinfo {pages}
  {035301} (\bibinfo {year} {2004})}\BibitemShut {NoStop}%
\bibitem [{\citenamefont {Lebedev}\ \emph {et~al.}(2010)\citenamefont
  {Lebedev}, \citenamefont {L'vov},\ and\ \citenamefont {Nazarenko}}]{LLN}%
  \BibitemOpen
  \bibfield  {author} {\bibinfo {author} {\bibfnamefont {V.~V.}\ \bibnamefont
  {Lebedev}}, \bibinfo {author} {\bibfnamefont {V.~S.}\ \bibnamefont {L'vov}},
  \ and\ \bibinfo {author} {\bibfnamefont {S.~V.}\ \bibnamefont {Nazarenko}},\
  }\href@noop {} {\bibfield  {journal} {\bibinfo  {journal} {J .Low Temp.
  Phys.}\ }\textbf {\bibinfo {volume} {161}},\ \bibinfo {pages} {606} (\bibinfo
  {year} {2010})}\BibitemShut {NoStop}%
\end{thebibliography}
%

\end{document}